\documentstyle[twocolumn,aps]{revtex}
\begin{document}
\title{New Mechanism of Quantum Oscillations in the Superconducting
Mixed State ($H_{c1}\ll B\ll H_{c2}$)}
\author{Lev P. Gor'kov$^{1,2}$ and J.R. Schrieffer$^1$}
\address{$^1$National High Magnetic Field Laboratory, Florida State
University, Tallahassee, FL 32310}
\address{$^2$L.D. Landau Institute for Theoretical Physics, Russian Academy
of Sciences, 117334 Moscow, Russia}
\date{Received $~~~~~~~~~~~~~~$}

\maketitle

\begin{abstract}
We argue that inhomogeneity inherent to the presence of periodic
supercurrents in the vortex lattice sorts excitations by energies into
the ones that are spatially localized and those that perform motion
along large Larmour orbits.  This energy threshold results in a new
mechanism for the de Haas-van Alphen oscillations which enhances
oscillations at $B\ll H_{c2}$, even for an isotropic superconductor with
a constant gap.  We suggest that the mechanism is of a general character
and can cause the slow decay of the de Haas-van Alphen effect when the
field, $B$, decreases below $H_{c2}$.
\end{abstract}

\vspace{.15in}

\pagestyle{empty}

\vspace{.25in}

The de Haas-van Alphen (dHvA) effect in the superconducting (SC)-state was
recently reported at magnetic fields, $B$, surprisingly smaller than the
upper critical field value, $H_{c2}$ [1,2].  Theoretically,
the dHvA-signal was expected to decay faster (see discussion and
references in [3]).  In the experiments [1,2] the effect remains
observable at $B$ as low as $0.3 \div 0.4H_{c2}$, where
the average SC-gap is thought to be too large for an electron
to perform a circular motion along orbits with the large Larmour radius,
$r_L\sim v_F/\omega_c$.  

Inhomogeneity of the current distribution and of the SC-order parameter
itself cause serious difficulties for a rigorous theoretical treatment.
Therefore in [3] an extreme limit of the lattice of isolated vortices,
\begin{eqnarray}
\xi_0\ll d\ll \delta_L~;~~H_{c1}\ll B\ll H_{c2}
\end{eqnarray}

\noindent has been chosen to find out whether any symmetry
zeroes in the gap function could enhance the
dHvA effect in the SC-state (in (1) $\xi_0$ is the coherence length,
$d$- the vortex lattice period, $\delta_L$- the penetration depth).

The results [3] confirmed expectations [4] that in the 3D-case existence
of a symmetry line of zeroes in the gap may restore the dHvA effect for
some field directions, although the effect is much weaker than predicted
in [4] due to scattering of electrons on flux lines.

As for a ``d-wave'' gap in layered superconductors, the levels'
systematics is such [3] that it forbids their crossing the chemical
potential with the field change -- the mechanism known to be the essence
of the dHvA effect in normal metals.  No qualitative difference is
expected, hence, between a ``d-wave'' and an ordinary anisotropic gap as
far as the dHvA effect is concerned.

Chances to observe the dHvA oscillations for a superconductor in
the regime (1) are rather vague, for the signal rapidly decreases
below the experimental resolution.  Common predictions for the signal's
amplitude would give an exponential factor, $\exp (-\Delta
/\omega_c)\ll 1$, where $\Delta$ is a gap scale and $\omega_c$ the
cyclotron frequency (e.g., see [5]).

On the theory part however, the limit (1) suggests significant
simplifications, because the spatial distribution of the field,
currents, and the gap are all well known from the phenomenological
consideration [6].  Main contributions are expected to come from the
``bulk'' since the volume occupied by vortex cores is small.

In that which follows the field range (1) is chosen again to {\it
demonstrate} the existence of some {\it new} quantum oscillations
mechanism which is specific to the SC-state and is originated by spatial
variation of the local density of states.  Although we expect that the
effect should have a more general character, we have chosen the regime
(1) and the isotropic gap for its simplicity.

In the SC-state an excitation bears electron- or hole-like features to
the extent its energy, $\varepsilon({\bf
p})=\sqrt{v^2_F(p-p_F)^2+\Delta^2}$, exceeds $\Delta$.  Taking
supercurrents into account, the local spectrum becomes
\begin{eqnarray}
E({\bf p})=\varepsilon({\bf p})+{\bf p}\cdot {\bf v}_s({\bf r})
\end{eqnarray}

\noindent where ${\bf v}_s({\bf r})$ stands for distribution of
superfluid velocity.  Hence, the second term in (2) generates a
potential relief, such that an excitation with a momentum ${\bf p}$ may,
or may not, perform infinite motion depending on whether its energy
exceeds, or not, the value $\Delta +max|{\bf p}\cdot {\bf v}_s({\bf
r})|$.  An excitation (2) experiences a Lorentz force on the part of
the magnetic field.  Whether an excitation may perform an itinerant
motion is equivalent to the condition that such excitations can encircle
a large Larmour orbit.  The additional potential in (2),
thus, imposes an energy boundary between ``localized'' and ``extended''
states.  The latter may contribute to magnetization.

To show that the above simple arguments, indeed, lead to a new
contribution to quantum oscillations in the SC-state, we explore
further the equations derived in [3] for a superconductor in a magnetic
field.  The method [3] consists of averaging of the Gor'kov
equations over classical trajectories in the magnetic field.

The core of the method is given by eqs. (24-27) of [3].  We use the same
notations [3] below and start with deriving the new Schrodinger equation
with the $h(\varphi)$-terms (eq. (31) [3])included and
$\Delta\equiv$const.:
\begin{eqnarray}
-\omega_c^2y^{''}-2Eh(\varphi)y=(E^2-\Delta^2)y
\end{eqnarray}

\noindent Note that the $h(\varphi)$-term of eq. (31) [3] is nothing but
the Doppler shift in (2).  In the derivation of (3) terms of the order
of $h^2(\varphi)$ and $\omega_c(dh(\varphi)/d\varphi)$ were omitted.  To
estimate the second term recall that the dependence $h(\varphi)$ on
$\varphi$ originates from the dependence of ${\bf v}_s({\bf r}(t))$ on
$t$: $t\rightarrow \varphi/\omega_c$.  An electron moving along the
orbit with a velocity $v_F$ senses changes in ${\bf v}_s({\bf r})$ with
a frequency $v_F/d$.  A change  on distances $\sim d$ is
therefore equivalent to its change on the scale, $\delta\varphi$, of the
order of
\begin{eqnarray}
\delta\varphi\sim (d\omega_c/v_F)\sim d/r_L
\end{eqnarray}

\noindent The $h(\varphi)$-term being $h(\varphi)\sim
v_F/d$, one obtains:
\begin{eqnarray}
h^2(\varphi)\sim\omega_ch'(\varphi) \sim v_F^2/d^2\sim
\Delta^2(\xi_0/d)^2 \nonumber
\end{eqnarray}
In (3) at $E\sim\Delta$ one has:
\begin{eqnarray}
Eh(\varphi)\sim\Delta^2(\xi_0/d)\ll\Delta^2
\end{eqnarray}

As it was argued in [3], the oscillatory part of the magnetization is
contained (eq. (6') [3]) in the expression:
\begin{eqnarray}
M=-\frac{\mu e}{2\pi
c}\sum_{\lambda,\sigma}\left[\overline{|u_{\lambda}(\varphi)^2|}
n(E_{\lambda})\right.
\end{eqnarray}

\noindent where the index $\lambda$ numerates the eigenvalues.  At
$T=0~~n(E_{\lambda})\equiv 1$ (all states below zero are occupied).  The
bar $\overline{(\ldots)}$ means the normalization
integral, $(2\pi)^{-1}\int_0^{2\pi}(\ldots)d\varphi$.  Oscillatory
effects, if any, come  from the vicinity of the
chemical potential [3]:
\begin{eqnarray}
\mu =\omega_cN_0+\bar{\mu}
\end{eqnarray}

\noindent The chemical potential being large, $\mu/\omega_c\gg 1$, the
oscillatory part of magnetization in (6) does not depend on a specific
$N_0$.  The periodic (in $B^{-1}$) pattern in $M$ of eq. (6) may,
therefore, be expressed as a function of
\begin{eqnarray}
\kappa=\bar{\mu}/\omega_c
\end{eqnarray}

\noindent with $\kappa$ varying in the interval (0,1).  The eigen
functions $y(\varphi)$ themselves are not periodic.  The periodicity
conditions are to be imposed on:
\begin{eqnarray}
y(\varphi)\exp\{-i\kappa\varphi\} 
\end{eqnarray}

\noindent With (8,9) in mind, the problem of solving eq.(3) becomes
equivalent to the one of finding the band structure of a particle moving
in the periodic potential, $2Eh(\varphi)$, where $\kappa$ (8) plays the
role of a quasimomentum, $\omega_c^2\Rightarrow (2m_{eff})^{-1}$ and
$E^2-\Delta^2$ being the ``energy''.

Summation over $\lambda$ in (6) is routinely replaced by the integration
by making use of the Poisson formula:
\begin{eqnarray}
\sum^{+\infty}_{n=-\infty}\delta(\lambda
-n)=\sum^{+\infty}_{K=-\infty}\exp (2\pi iK\lambda)
\end{eqnarray}

\noindent Applying (10) to eq. (6), integrating over $\lambda$ by parts,
one would arrive at terms of the form:
\begin{eqnarray}
M_{osc}=\frac{2i\mu
e}{(2\pi)^2c}\sum_K\frac{1}{K}\int^{+\infty}_{-\infty}
e^{2i\pi K\lambda}
\frac{d}{d\lambda}\left(\overline{|u_{\lambda}(\varphi)|^2}\right)
d\lambda
\end{eqnarray}

\noindent first noticed in [3].
Eqs. (10) and (11) have no value unless a relation between $\lambda$ and
energy is established.  Below we will construct a function
\begin{eqnarray}
\lambda(E)=\Phi(E)/2\pi
\end{eqnarray}

\noindent such that all the eigenvalues in (6) are determined as in
(10), by the provision:
\begin{eqnarray}
\lambda(E_{\lambda})=n
\end{eqnarray}

\noindent If $\Phi(E)$ were known on the real $E$-axis, it then may be
analytically continued onto the complex plane.  

With this in mind first simplify
notation in (3)
\begin{eqnarray}
-\varepsilon =E+\Delta -h_{max}(\varphi_o)
\end{eqnarray}

\noindent $h_{max}(\varphi_0)$ is the maximum of $h(\varphi)$ along
a given trajectory.  One may assume $\varphi_0=0$.  Eq. (3) becomes:
\begin{eqnarray}
-\omega_c^2y''+2\Delta(h(\varphi)-h_{max})y=2(-\varepsilon)\Delta y
\eqnum{3'}
\end{eqnarray}

\noindent Even though $h\sim v_F/d\sim\Delta (\xi_0/d)\ll\Delta$, the
ratio
\begin{eqnarray}
\Delta h/\omega_c^2\gg 1
\end{eqnarray}

\noindent is large. In the WKB-approach:
\begin{eqnarray}
y(\varphi)=ay_{+}(\varphi)+by_-(\varphi)
\end{eqnarray}

\noindent where $y_{\pm}(\varphi)$ are of the form [7]:
\begin{eqnarray}
y_{\pm}(\varphi)=A(S')^{-1/4}\exp \left(\pm
i\int_0^{\varphi}S'(\varphi)\right)
\eqnum{16'}
\end{eqnarray}

\noindent Here $A$ is the normalization factor and 
\begin{eqnarray}
S'(\varphi)=\left(\sqrt{2\Delta}/\omega_c\right)
\sqrt{h_{max}-h(\varphi)-\varepsilon}
\end{eqnarray}

\noindent At $(-\varepsilon)>0$ the quantization condition (taking (9)
into account) gives:
\begin{eqnarray}
S(2\pi ,-\varepsilon_n)=\frac{\sqrt{2\Delta}}{\omega_c}\int^{2\pi}_0
\sqrt{h_{max}-h(\varphi)-\varepsilon_n} \nonumber \\ 
=2\pi n+2\pi\kappa
\end{eqnarray}
Eq. (18) at $|\varepsilon |\gg\Delta$ matches the spectrum of free
electrons in the magnetic field (in presence of the flux currents).  At
$(-\varepsilon)<0$ one would obtain ``localized'' states.  Let us
justify first this last statement.

The ``potential,'' $h(\varphi)-h_{max}<0$ in (17) is rather irregular
and has many minima alternating with local maxima on
distances of order of $\delta\varphi$, (4).  The probability to tunnel
between minima is
measured by the value of $\exp (-|\delta S|)$ with $\delta S
\sim\left(d/\xi_0\right)^{1/2}\gg 1$,  i.e. is small enough to
consider barriers as impenetrable ones.

The two WKB-branches, one at $(-\varepsilon) >0$ and the other for
$(-\varepsilon)<0$, cannot together form a
function $\Phi(E)$ such as eq. (12).  The WKB-approach is invalid at
$(-\varepsilon)$ too close to the maximum of $h(\varphi)$.  Indeed,
expanding (18) at small $(-\varepsilon)>0$ results in the singularity:
\begin{eqnarray}
S(2\pi ,-\varepsilon) \simeq S(2\pi ,0)+\frac{(-\varepsilon)}{2\omega_c
a^{1/2}}\left[\ln \left(\frac{\Delta
a(\delta\varphi)^2}{-\varepsilon}\right)\right]
\end{eqnarray}
For $h(\varphi)$ close to the maximum we have chosen:
\begin{eqnarray}
h(\varphi)=h_{max}-(a\Delta\varphi^2/2)
\end{eqnarray}
with $ a\Delta\sim v_F/d(\delta\varphi)^2$.

Thus, in the WKB-approximation there are two sorts of solutions:  those
of the form of eq. (16) at $(-\varepsilon)>0$, and the ``localized''
states at $(-\varepsilon)<0$.  It is useful to show how these findings
are connected with our intuitive expectations regarding the role of the
Doppler term (2).

The solutions $u_{\lambda}(\varphi)$ and $v_{\lambda}(\varphi)$ must be
normalized together: $\overline{|u_{\lambda}(\varphi)|^2}+
\overline{|v_{\lambda} (\varphi)|^2}=1$.  There are two useful auxiliary
relations:
\begin{eqnarray}
\overline{|u^2|}=(1/2)\left\{
\overline{|y|^2}+(i\omega_c/2E)\overline{(y^{\ast}y' -yy^{\ast
'})}\right\} \nonumber\\
\overline{|v^2|}=(1/2)\left\{
\overline{|y|^2}-(i\omega_c/2E)\overline{(y^{\ast}y' -yy^{\ast
'})}\right\}
\end{eqnarray}
which may be derived with the same accuracy as (3).  One concludes from
(21) that $\overline{|y|^2}=1$, and
$\overline{|u_{\lambda}|^2}=\frac{1}{2}$ {\it independently on the
energy for all localized states} for which the wave functions are real.
Therefore there is, indeed, a threshold singularity in $M_{osc}$ (11)--
``localized'' states give no contribution into $M_{osc}$.  To study the
phenomenon rigorously one must go beyond the WKB-accuracy.

 At the energy close to the maximum of the potential in eq. (3')
one may apply the WKB solutions (16, 16') only far away from points
$\varphi =0$ and $\varphi =2\pi$.  In the vicinity of $\varphi =0$ eq.
(3'), (20) is solvable in terms of the
parabolic cylinder functions.  Matching the asymptotic behavior of the
latter to the WKB-form (16, 16') makes it possible to find
relations between coefficients $(a,b)$ in (16) to the right of
$\varphi=0$, and the similar set $(a',b')$, to the left of it: 
\begin{eqnarray}
\left(\begin{array}{c} a'\\ b' \end{array}\right)
=\left(\begin{array}{cc}
\alpha & \beta \\ \beta^{\ast} & \alpha^{\ast} \end{array}\right)
\left(\begin{array}{c} a\\ b \end{array}\right) ~;~~\left(|\alpha
|^2-|\beta |^2=1\right)
\end{eqnarray}
Moving along $\varphi$ from $\varphi =(0)_+$ and using (16), one reaches
close to the point $\varphi =(2\pi)_-$.  With the help of (22), the
solution then may be transformed into the solution at $\varphi
=(2\pi)_+$.

The periodicity (see (9)) leads to the equation:
\begin{eqnarray}
R(l)\equiv |\alpha |\left(e^{+i\tilde{S}}+e^{-i\tilde{S}}\right) =2\cos
2\pi\kappa
\end{eqnarray}
Here and below the dimensionless energy, $l$, is
\begin{eqnarray}
l=2(-\varepsilon)/\omega_c a^{1/2}
\end{eqnarray}
In (23) we used notations:
\begin{eqnarray}
\alpha =|\alpha |\exp (i\theta)~;~~\tilde{S}=S(2\pi ,l)-\theta(l)
\end{eqnarray}
($S(2\pi ,l)$ coincides with (19) re-written in terms of $l$ from (24)).
Of the two solutions in eq. (23) we have to take 
\begin{eqnarray}
e^{i\tilde{S}}=\frac{\cos 2\pi\kappa}{|\alpha |}+\sqrt{\left(\frac{\cos
2\pi\kappa}{|\alpha |}\right)^2-1}\equiv \rho
\end{eqnarray}
Indeed, at $l\rightarrow +\infty$,   $\alpha\Rightarrow 1$,
eq. (25) for energy levels goes over into the WKB-result (18).

Introduce the function
\begin{eqnarray}
\Phi(-\varepsilon)\equiv \Phi (l)=S(2\pi
,l)-\theta(l)-\frac{1}{i}\ln\rho (l)
\end{eqnarray}
As we just mentioned, at large $l$ solutions of $\Phi(l_n)=2\pi n$ go
over into the WKB-spectrum which at very large energies, $l\gg 1$,
transforms into the spectrum of free electrons in the magnetic field.
$\Phi(l)$ is a ``monotonous'' real function at $l\sim 1$.  At $l<0$ eq. (13) 
also determines the WKB-localized levels immediately below $h_{max}$.
Therefore we may use (27) as the definition of $\lambda (E)$ (12),
{\it including} energies in the vicinity of $h_{max}$ where our phenomenon
takes place.  Correspondingly, the integral in (11) over $\lambda$
acquires the meaning of the integration over energy:
$d\lambda\Rightarrow (d\lambda/dE) dE \equiv \left( d\lambda /dl\right) dl$.

Consider (27) in some more details.  The transition coefficient,
$|\alpha |$, on the real axis is of the form: 
\begin{eqnarray}
|\alpha |=(1+e^{-\pi l})^{1/2}
\end{eqnarray}
It has the branch-cuts at points $l_m=\pm(2m+1)i~(m=0,1,\ldots)$ which
should be chosen along {\it imaginary} axis, for $|\alpha |$ to be the single
valued analytical function inside a strip near the real axis.  The
phase, $\theta(l)$, displays similar properties at $|l|\sim 1$ (see $\tilde{S}(l)$ (31)).  As for $\rho (l)$ (26),
the square root in (26) has singularities at:
\begin{eqnarray}
l'_m=\pm (2m+1)i-\frac{1}{\pi}\ln (\sin^22\pi \kappa)
\end{eqnarray}
Bending the contour of integration over $l$ into the upper half-plane,
the main contribution into $M_{osc}$ (11) is to be due to ``nearest''
singularities, at $l_m$ and $l_m'$.

These values have the scale $|\lambda |\sim 1$, which, being
re-written in terms of the energy, (24), results in:
\begin{eqnarray}
|\varepsilon |\sim\omega_c a^{1/2}\sim h(\xi_0/d)^{1/2}\ll h
\end{eqnarray}
Eq. (30) justifies our implicit assumptions above that the effect is
caused by the main peak, $h_{max}$. Other peaks,
although being of the same scale, $h\sim v_F/d$, do not contribute to
the effect produced by the fine structure (30) of the levels developing
in the very vicinity of $h_{max}$.

The subsequent analysis although simple and straightforward, involves
somewhat lengthy mathematical  details to be published
elsewhere.  Below we provide only a sketch of a few major steps, before
discussing the result.

Exact $\Phi(l)$, eq. (27), has no 
non-analytic  $l\ln l$-term in (19) at small $l$.  At $l\sim 1$ one
gets $(c\sim 1)$:
\begin{eqnarray}
\tilde{S}(l)=S(2\pi
,0)+\frac{1}{2}l\left[\ln\left(\sqrt{\frac{d}{\xi_0}}\right) +c\right]
-\nonumber \\
-\frac{1}{2i}\ln\left[\frac{\Gamma(\frac{1}{2}+\frac{il}{2})}{\Gamma(\frac{1}{2}-\frac{il}{2})}\right]
\end{eqnarray} 

The important step is to calculate $\overline{|u_l|^2}$ in (11).  In
particular, the normalization is to be known with better accuracy than
provided by the WKB-expressions of eq. (16, 16').  This can be achieved
in frameworks of the above method.  Fortunately, however, properties of
the Bloch functions for a one-dimensional periodical potential are well
studied.  With the help of eq. (4.18) in ref. [8] and our eqs. (21) we
derived:
\begin{eqnarray}
\overline{|u_l|^2}=\frac{1}{2}-\pi a^{1/2}\sin 2\pi\kappa
\left[\frac{dR}{dl}\right]^{-1} 
\end{eqnarray}
where $R(l)$ is the R.H.S. of eq. (23).  Expression for $M_{osc}$ (11)
is given as a sum of integrals, $I_K$, which may be re-written as
integrals over the energy variable, $l$:
\begin{eqnarray}
I_K=\int_{-\infty}^{+\infty}
e^{iK\Phi(l)}\frac{d}{dl}\left(\overline{|u_l|^2}\right) dl
\end{eqnarray}
The constant, i.e. $1/2$, having been  eliminated in (33),  the path of
integration over real $l$ can be transformed into two contours,
$C_1$ and $C_2$, encircling the two branch-cuts caused by  singular
points of $\Phi(l), ~l_m$, and $l_m'$ of eq. (28, 29).  With
integrations over $C_1$ and $C_2$ now running parallel to the imaginary axis
integrals become rapidly (exponentially) convergent, and integration
back by parts is allowed.  One obtains two integrals of the form:
\begin{eqnarray}
I_K=\frac{iK\pi a^{1/2}\sin 2\pi\kappa}{2} \int \frac{\exp
iK\Phi(l)dl}{(\sin^22\pi\kappa +e^{-\pi l})^{1/2}} 
\end{eqnarray}
With $\Phi(l)$ given by (27) and (31), and the proper definition of the
branches on the cuts,  integrations can be completed
analytically, particularly so in the limit
$\ln\left(\sqrt{d/\xi_0}\right)\gg 1$.  Without providing here
the result explicitly, we use this latter assumption to discuss the order of
magnitude for $I_{K=1}$:
\begin{eqnarray}
I_{K=1}\propto
a^{1/2}e^{iS(2\pi,0)}\left(\xi_0/d\right)^{1/4}\sim \nonumber\\
\sim\Delta /\omega_c\left(\xi_o/d\right)^{7/4}e^{iS(2\pi,0)}
\end{eqnarray}
The numerical coefficient in (35) is large but there is a phase factor
containing 
\begin{eqnarray}
S(2\pi,0)=\sqrt{2\Delta}/\omega_c\int^{2\pi}_0
(h_{max}-h(\varphi))^{1/2}d\varphi
\end{eqnarray}
and the result has to be averaged yet over all trajectories.

Our estimates proceed in the following way.  Since the Larmor orbit is
so large $(r_L\gg d)$, the maximum in $h(\varphi)$ must be practically
the same for any trajectory, while its position, $\varphi_0$, varies.
The position, $\varphi_0$, played no role in our analysis above.  The
integral in (36) is positive, therefore we write
\begin{eqnarray}
e^{iS(2\pi,0)}=e^{i\overline{S(2\pi,0)}}e^{i\delta S} \nonumber
\end{eqnarray}
where $\overline{S(2\pi,0)}$ is an average, i.e. {\it just a phase}
(although unknown), while $\delta S$ fluctuates around
$\overline{S(2\pi,0)}$.  Since
$\overline{S(2\pi,0)}\sim\Delta /\omega_c
\left(\xi_0/d\right)^{1/2}$ is large, $\langle e^{i\delta
S}\rangle$ can be calculated using for $\delta S$ the Gaussian
distribution with $\overline{\delta S^2}=\overline{S(2\pi,0)}$.  For
$M_{osc}^S/M_{osc}^N$ it gives:
\begin{eqnarray}
\sim
\Delta /\omega_c\left(\xi_0/d\right)^{7/4} \exp \left[
-\Delta /\omega_c\left(\xi_0/d\right)^{1/2}\right]
\end{eqnarray}

The exponent (37) remains, of course, yet too small to
expect that the dHvA effect may be measured experimentally in the regime
(1).  Nevertheless, we have demonstrated the effect of the local density
of states characteristic of the superconducting state: an effective
Dingle's temperature becomes reduced as 
$\Delta\Rightarrow\Delta(\xi_0/d)^{1/2}$ even for the isotropic BCS-like
superconductor.  In the anisotropic case such a reduction is expected to
be stronger [9] $\Delta\Rightarrow\Delta(\xi_0/d)$,
in the same regime (1).

To conclude, we demonstrated that, unlike normal metals where the
dHvA-effect is caused by electron levels crossing the chemical
potential, in the superconducting heterogeneous state oscillations are
caused by levels crossing the energy threshold separating ``localized''
and itinerant states.  We argue that this mechanism may be responsible
for pronounced oscillations in the regime $B\sim H_{c2}$ where not only
is the gap weaker, but is heterogeneous (periodic) itself, together with
distributions of the field and currents. 

This work was supported by the NHMFL through NSF cooperative agreement
No. DMR-9016241 and the State of Florida and NSF grant DMR-96-29987 (JRS).

\vspace{.25in}

\noindent {\bf References}

\vspace{.25in}

\noindent [1] T. Terashima, C. Haworth, H. Takeya, S. Uji, H. Aoki, \\
\indent $~$and
K. Kadowaki, Phys. Rev. {\bf B56}, 5120 (1997).

\noindent [2] Y. Onuki, to be published in Proceedings of the SCES \\
\indent $~$'98,
Paris, 1998.

\noindent [3] L.P. Gor'kov and J.R. Schrieffer, Phys. Rev. Lett. \\
\indent $~${\bf 80}, 3360 (1998).

\noindent [4] K. Miyake, Physica (Amsterdam) {\bf 186B}, 115 (1993).

\noindent [5] K. Maki, Phys. Rev. {\bf B44}, 2861 (1991); M.J. Stephen,
\\ \indent $~$Phys. Rev. {\bf B45}, 5481 (1992).

\noindent [6] A.A. Abrikosov, Sov. Phys. JETP {\bf 5}, 1174 (1957).

\noindent [7] L.D. Landau and E.M. Lifshitz, {\it Quantum Mechanics:}
\\ \indent $~${\it nonrelativistic theory} (Pergamon Press, New York, \\
\indent $~$1977).

\noindent [8] W. Kohn, Phys. Rev. 115, 809 (1959).

\noindent [9] L.P. Gor'kov, submitted to JETP Lett., 1998.
        
\end{document}